\pdfoutput=1
\documentclass[conference, a4paper, 10pt, twoside]{IEEEtran}
\usepackage[english]{babel}
\usepackage[T1]{url}
\usepackage[pdftex]{graphicx}
\usepackage[svgnames]{xcolor}
\usepackage[pdftex, unicode, bookmarks=true, bookmarksnumbered=true,
  pdfpagemode={UseOutlines}, plainpages=false, pdfpagelabels=true,
  colorlinks=true, linkcolor={DarkBlue}, citecolor={DarkBlue}, 
  urlcolor={DarkBlue}]{hyperref}
\usepackage{enumitem}
\usepackage[final]{microtype}
\usepackage[utf8]{inputenc}
\usepackage[LGR,T1]{autofe}

\newcommand{\mailto}[1]{\href{mailto:#1}{\nolinkurl{#1}}}
\newcommand{\doi}[1]{\href{http://dx.doi.org/#1}{\nolinkurl{#1}}}

\makeatletter
\let\thanks\@IEEESAVECMDthanks
\makeatother

\title{A CMOS Tailed Tent Map for the Generation of
  Uniformly Distributed Chaotic Sequences} 

\author{%
  \parbox{0.47\linewidth}{\begin{center}%
      Sergio~Callegari, Gianluca~Setti\\[1ex]
      DEIS, University of Bologna, Italy\\
      \mailto{scallegari@deis.unibo.it}\\
      \mailto{gsetti@deis.unibo.it}\end{center}}
  \hfill
  \parbox{0.47\linewidth}{\begin{center}%
      Peter~J.~Langlois\\[1ex]
      EEEng Dept., King's College, London, UK\\
      \mailto{p.langlois@bay.cc.kcl.ac.uk}\\
      \hspace*{0pt}\end{center}}%
  \thanks{This is a post-print version of a paper presented at the 1997
    IEEE International Symposium on Circuits and Systems (ISCAS '97).
    Published paper available via DOI
    \doi{10.1109/ISCAS.1997.621829}. Cite
    as:\protect\\[1ex]
    S.~Callegari, G.~Setti, P.~J.~Langlois ``A CMOS Tailed Tent Map for
    the Generation of Uniformly Distributed Chaotic Sequences'',
    \emph{Proc.\@ of 1997 IEEE International Symposium on Circuits and
      Systems, (ISCAS '97)}, vol.~2, pp.~781--784, Jun.\@ 1997
    \protect\\[1ex]
    Copyright © 1997 IEEE. Personal use of this material is
    permitted. However, permission to use this material for any other
    purposes must be obtained from the IEEE by sending a request to
    \mailto{pubs-permissions@ieee.org}.\protect\\[-2ex]}}
  
\hypersetup{%
  pdftitle={%
    A CMOS Tailed Tent Map for the Generation of
    Uniformly Distributed Chaotic Sequences}, 
  pdfauthor={%
    Sergio Callegari,
    Gianluca Setti,
    Peter J.~Langlois}}

\renewcommand\normalsize{\fontsize{9.6}{11.52}\selectfont}

\begin{document}
\maketitle
\thispagestyle{empty}
\begin{abstract}
This paper describes the design of a modified tent map characterized
by a uniform probability density function.  The use of this map is
proposed as an alternative to the tent map and the Bernoulli
shift. It is shown that practical circuits
implementing the latter two maps may possess parasitic stable
equilibria, fact which would prevent the desired chaotic behavior of
the system. On the other hand, commonly used strategies to avoid the
parasitic equilibria onset also affect the uniformity of the
probability density function. Conversely, the use of the proposed
tailed tent map allows to assure a certain degree of parameter
deviation robustness, without compromising on the statistical
properties of the system.
\end{abstract}

\section{Introduction}
Discrete time chaotic systems based on simple, one-dimensional, piecewise
linear maps can be conveniently implemented using current mode
operation~\cite{Bean:ISCAS94-6-125}. Low circuit complexity makes them
appealing for many applications, like secure communication
schemes~\cite{Hasler:ISCAS94}, noise generators and analog random number
generation for stochastic neural models~\cite{Clarkson:NNW93-5-551}.
Design requirements can be posed either on the spectrum or on the
probability distribution of the chaotic signal. For instance, the random
number generator needed in~\cite{Clarkson:NNW93-5-551} should produce
samples characterized by a uniform probability density function (PDF).

\begin{figure}[ht]
\begin{center}
\begin{minipage}{0.4\columnwidth}
\resizebox{\linewidth}{!}{\includegraphics{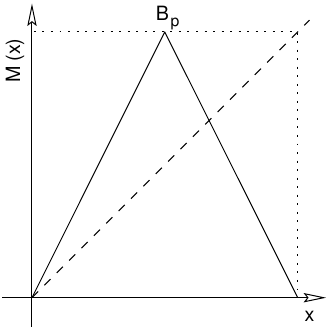}}\par
\centerline{\sffamily\small(A)}
\end{minipage}
\hspace{0.2cm}
\begin{minipage}{0.4\columnwidth}
\resizebox{\linewidth}{!}{\includegraphics{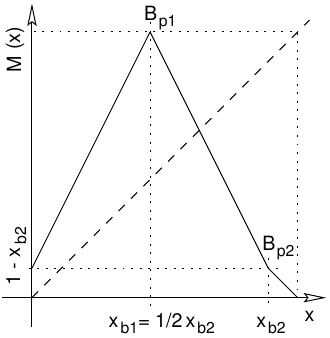}}\par
\centerline{\sffamily\small(B)}
\end{minipage}
\end{center}
\caption{\label{fig:maps}The tent map (A) and
the tailed tent map (B). Normalized graphs in $I=[0,1]$.}
\end{figure}

Although systems based on the tent map (Fig.~\ref{fig:maps}A) and on
the Bernoulli shift satisfy this requirement, their
implementation \cite{Bean:ISCAS94-6-125,Delgado:El} is not completely
straightforward, due to the possible onset of parasitic stable
equilibrium points, as result of the unavoidable errors introduced
by the physical realization of the electronic devices. Since this fact
would not allow the desired chaotic behavior, special strategies must
be adopted to make the system robust to parameter
deviation. Unfortunately, commonly adopted solutions~\cite{Delgado:El}
also prevent the achievement of a uniform PDF.
 
The aim of this paper is to show that a one-dimensional system based on a
modified version of the tent map, referred as \emph{tailed tent map}
(Fig.~\ref{fig:maps}B), possesses both a uniform PDF and a
chaotic behavior inherently robust to implementation errors, being
better suited for carrying out, for instance, the  random number
generator needed in~\cite{Clarkson:NNW93-5-551}.
In order to show the feasibility of the system, a design in a standard
$1\mu m$,  n-well, single-poly CMOS technology is also proposed. 
Several circuit  simulations confirm the uniform distribution of
the samples produced by the system.

\subsection{Mathematical background}
Let $I=[a,b]$ be an interval of the real line ${\mathcal R}$ and
consider the one-dimensional dynamical system 
\begin{equation}
\label{1Dsys}
x_{n+1} = M(x_n) 
\end{equation}
where $M$ is a nonsingular transformation of $I$ into itself. When
designing a pseudo random number or a noise generator based on
system~(\ref{1Dsys}) two properties are particularly important, namely
\emph{i}) chaotic behavior and \emph{ii}) uniform probability density.
\begin{enumerate}[wide, label={\roman*)}]
\item Chaotic behavior is assured when the system
(\ref{1Dsys}) has a positive Lyapunov exponent, defined as
\cite{Ott:CDS}:
\begin{equation}
\label{eq:chaoticity}
h=\lim_{T\rightarrow\infty}\frac{1}{T}\sum_{i=0}^{T-1}\ln|M'(x_i)|
\end{equation}
Roughly speaking, the Lyapunov exponent is an index of the
sensitivity of the system to initial conditions and therefore of its level of
\emph{unpredictability}. From (\ref{eq:chaoticity}) it immediately
follows that  systems based on everywhere expanding maps (for which
$\mbox{inf}_{x \in I}|M'(x)| >1$) are chaotic.  
\item Consider an interval $S \subseteq I$ and let 
$\chi_{S}$ be the relative characteristic function. Then, the
``average time'' spent in  $S$  by an orbit originating in
$x_0$ may be defined as
\begin{equation}
\label{eq:averagetime}
\lim_{n\rightarrow\infty}\frac{1}{n}
\sum_{i=0}^{n-1}\chi_S(M^i(x_0)),
\end{equation}
where $M^{i}$ is the $i$-th iterate of the map.
Note that if limit (\ref{eq:averagetime}) exists, it depends on the
choice of the initial condition $x_0$. Birkhoff ergodic theorem
\cite{Lasota:CFN} gives a sufficient condition for the existence of
limit (\ref{eq:averagetime}) ``almost'' independently from $x_0$. In
particular, it states that, if $M$ is \emph{measure
preserving}\footnote{$M$ is said to be measure preserving if a 
measure $\mu$  exists such that $\mu(M^{-1}(A))=\mu(A)$ for any measurable set
$A\subseteq I$.} and \emph{ergodic} then there is a \emph{unique} PDF
$\rho$ for $M$ such that
\begin{equation}
\label{eq:time=stat}
\lim_{n\rightarrow\infty}\frac{1}{n}
\sum_{i=0}^{n-1}\chi_S(M^i(x_0))=\int_{S}\rho(\xi)d\xi \ \ \ \mbox{a.e.} 
\end{equation}
for any measurable $S \subseteq I$.  Basically, ergodicity provides that
any initial condition randomly chosen in $I$ leads to a trajectory
characterized by the same statistical properties.  Furthermore, ergodicity
enables extraction of information about time-based statistics from the a
priori knowledge of the PDF.  In particular, it can be shown that, if $M$
is ergodic, the PDF $\rho$ is the unique \emph{invariant} under the
\emph{Perron-Frobenius} operator~\cite{Lasota:CFN} ${\mathcal
  P}_{M}:L^1\rightarrow L^1$ defined as
\begin{equation}
\int_{S}{\mathcal P}_{M}[\rho](\xi)d\xi = 
\int_{M^{-1}(S)} \rho(\xi)\,d\xi \ \ \ 
\forall S \subseteq I,
\end{equation} 
namely ${\mathcal P}_{M}[\rho]=\rho$.

In the following, we will consider the design of the tailed tent map
in order to obtain an ergodic system and we will thus determine its
uniform PDF as a fixed point of ${\mathcal P}_{M}$. Eventually,  we will
verify its robustness to breakpoints misplacement both via analytic
considerations and by approximating $\rho$ using time series data
extracted from simulations.
\end{enumerate}

\section{Systems based on the tent map}
Consider a tent map based system (\ref{1Dsys}), namely assume
$I=[0,1]$ and
\begin{equation}
\label{eq:tentmap}
M(x)=1-2|x-1/2|.
\end{equation}
Since the map is everywhere expanding, the system is chaotic. Moreover,
it can be proved that the system is ergodic and characterized
by a uniform PDF~\cite{Ott:CDS}. 

\begin{figure}[ht]
\centerline{
\vspace{1ex}
\resizebox{0.6\columnwidth}{!}{
\scalebox{1}[1]{\includegraphics{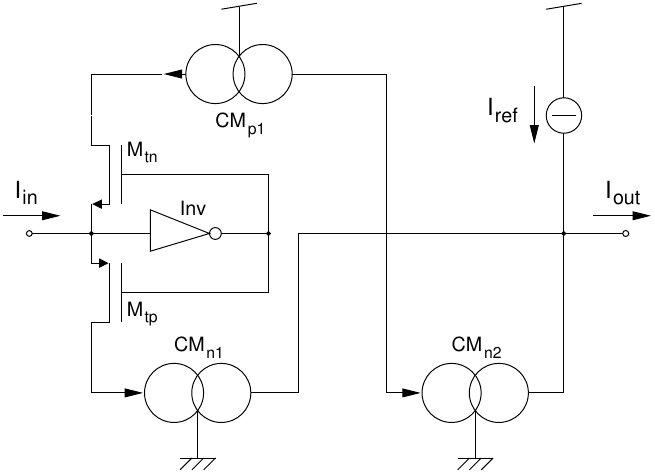}}}
}
\caption{\label{fig:tm-circuit} A circuit for the tent map.}
\end{figure}

Fig.~\ref{fig:tm-circuit} shows a suitable circuit implementation for
the map~\cite{Delgado:El}
Transistors $M_{tp}$, $M_{tn}$ and Inv act as an active rectifier: when
$I_{in}$ is positive (negative) $M_{tp}$ ($M_{tn}$) is on and $M_{tn}$
($M_{tp}$) is off. Since current mirrors $CM_{n1}$ and $CM_{p1}$ have a
2:1 ratio, one obtains $I_{out}=I_{ref}-2|I_{in}|$.  The simulated static
characteristic is represented in Fig.~\ref{fig:phaselock}A and is the same
as (\ref{eq:tentmap}) apart from an uninfluential change of the reference
system.  Considering the implementation of equation (\ref{1Dsys}) by a
current mode circuit, the necessary delay operation can be realized by
cascade connecting two switched current (SI) sample and hold stages with
complementary phase clock~\cite{Bean:ISCAS94-6-125}.

\subsection{Parasitic stable equilibrium and strategies to avoid it}
\begin{figure}[ht]
\begin{minipage}{0.45\columnwidth}
\begin{center}
\resizebox{\linewidth}{!}{%
\scalebox{1}[1.3]{\includegraphics{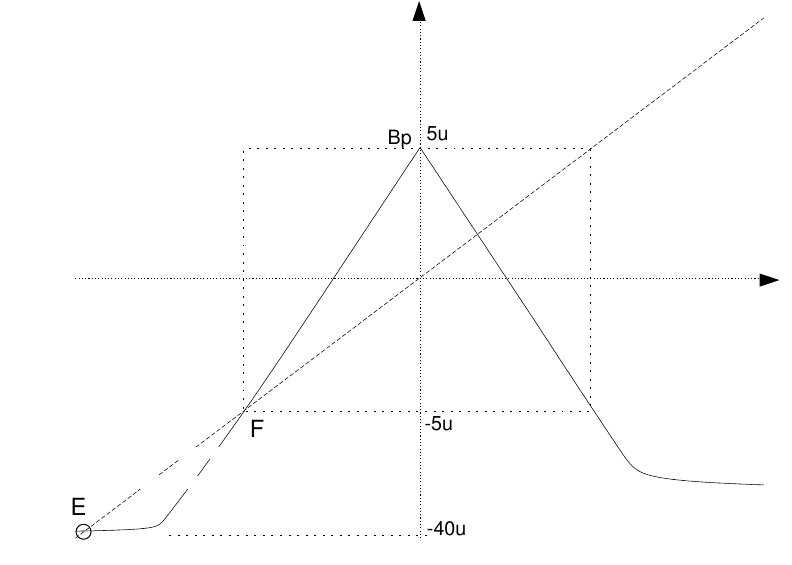}}}\par
\centerline{\sffamily\small(A)}
\end{center}
\end{minipage}
\hspace{0.2cm}
\begin{minipage}{0.45\columnwidth}
\begin{center}
\resizebox{\linewidth}{!}{\includegraphics{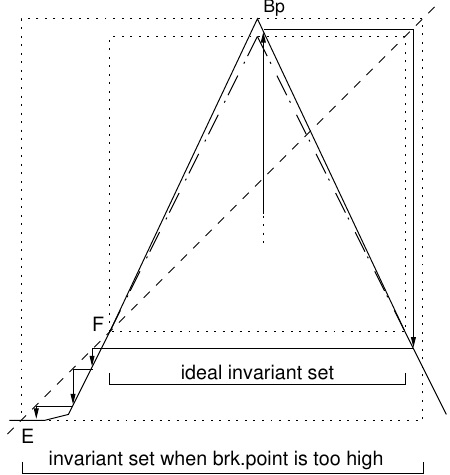}}\par
\centerline{\sffamily\small(B)}
\end{center}
\end{minipage}
\caption{\label{fig:phaselock}(A): Existence  of a parasitic stable
equilibrium  and (B): convergence of the system trajectory toward it,
due to a misplacement of $B_P$ in the map.}
\end{figure}

When the map is implemented using an analog circuit, some misplacement of
the breakpoint $B_P$ is unavoidable. If the ordinate of the breakpoint
exceeds its assigned value due to a slope error, a parasitic stable
equilibrium is introduced. In order to see how this happens, it is
necessary to consider the circuit static characteristic outside the map
domain. Due to the fact that the current mirrors or the current rectifier
transistors (Fig.~\ref{fig:tm-circuit}) reach their maximum allowable
current, the map slope decreases and flattens, as shown in
Fig.~\ref{fig:phaselock}A.  As a consequence a parasitic equilibrium point
exists (point E in Fig.~\ref{fig:phaselock}) characterized by $M'(x_E)<1$,
i.e.\ the equilibrium is stable.
Although this equilibrium always exists, it ideally
lies out of the map invariant set. On the contrary,
if the breakpoint ordinate is greater than its assigned
value (independently of the magnitude of the misplacement), the
map invariant set  abruptly expands, including also point $E$.
Therefore, after a transient phase, the system trajectory will
converge towards it, as schematically represented in Fig.~\ref{fig:phaselock}B.
The abrupt expansion of the invariant set
is due to the fact that its closure contains the unstable equilibrium
point $F$. As soon as the invariant set grows to include $F$ in its
interior it must also contain the whole interval $[x_{E},x_{F}]$.

Note that besides breakpoint misplacement, also noise added to the system
state may allow the system trajectory to converge to E.

\begin{figure}[ht]
\begin{center}
\begin{minipage}{0.4\columnwidth}
\resizebox{\linewidth}{!}{\includegraphics{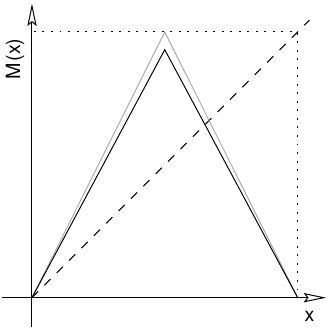}}\par
\vspace{-2ex}\centerline{\sffamily\small(A)}
\end{minipage}
\hspace{0.2cm}
\begin{minipage}{0.4\columnwidth}
\resizebox{\linewidth}{!}{\includegraphics{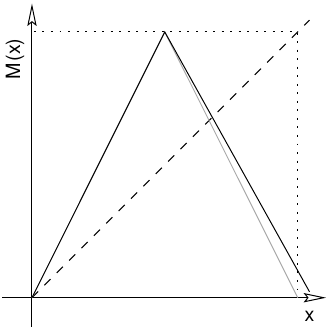}}\par
\vspace{-2ex}\centerline{\sffamily\small(B)}
\end{minipage}
\end{center}
\caption{\label{fig:strategies}Two strategies to avoid convergence to the
parasitic equilibrium. (A) shows a strategy which affects the
slope of both segments of the tent map, while (B) shows a strategy
which affects the slope of the second segment only. By rescaling the
graph it can be seen that also the case (B) is equivalent to lowering the
breakpoint.}
\end{figure}

In order to avoid this problem, several strategies can be used. All of
them are based on small modifications of the map\footnote{ Without loss of
  generality we will refer our considerations to normalized maps defined
  in $I=[0,1]$.}  and are equivalent to designing a system whose
breakpoint ordinate is less than 1, to assure an adequate margin. Two
typical solutions are shown in Fig.~\ref{fig:strategies}.

The margin one must allow  mainly depends on three factors:
\begin{enumerate}
\item the maximum random error one can expect for the circuit static
characteristic. This originates mainly from transistors mismatching;
\item the error introduced by the two SI sample and hold performing the
delay operation. This is mainly due to clock
feedthrough and appears as an undesired signal added to the
system state;
\item the error due to the dynamic response of the current comparator
used to implement the breakpoint (Fig.~\ref{fig:tm-circuit}). This 
error manifests itself as a sort of hysteresis and appears
when the circuit has to switch immediately over or below threshold. 
\end{enumerate}

In particular, if speed is an important requirement, transistors
dimensions must be reduced in order to contain parasitic capacitances
effects and the first factor will have great importance. On the other
hand, if a low-power circuit is desired, one has to accept a worsening of the
current comparator dynamic response and a consequent increased
relevance of the third factor. Therefore, particularly in fast, low
power systems the required margin can be appreciable.

\subsection{Pitfalls in the strategies to prevent convergence to
parasitic equilibria} 
\begin{figure}[ht]
\vspace{1ex}
\begin{minipage}{0.45\columnwidth}
\begin{center}
\resizebox{\linewidth}{0.8\linewidth}{\includegraphics{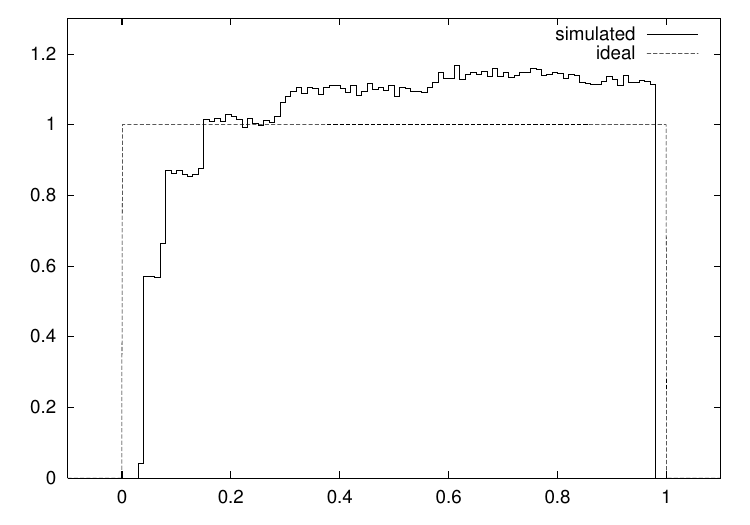}}\par
\vspace{-0.8ex}\centerline{\sffamily\small(A)}
\end{center}
\end{minipage}
\hspace{0.2cm}
\begin{minipage}{0.45\columnwidth}
\begin{center}
\resizebox{\linewidth}{0.8\linewidth}{\includegraphics{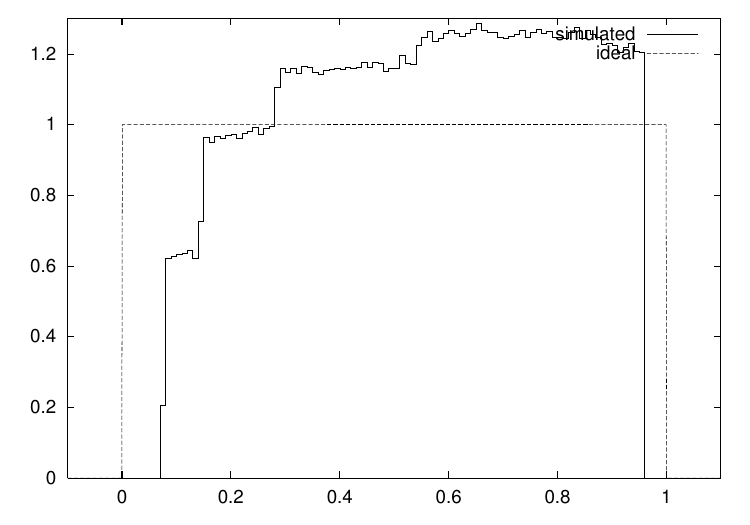}}\par
\vspace{-0.8ex}\centerline{\sffamily\small(B)}
\end{center}
\end{minipage}
\caption{\label{fig:pdf-tm}Consequences on the PDF of the strategy
shown in Fig.~\protect{\ref{fig:strategies}}A}
\end{figure}

Adopting one of the above mentioned strategies 
implies sacrificing  PDF uniformity. For instance, Fig.~\ref{fig:pdf-tm}
shows the consequences on the PDF of modifying the map as shown
in Fig.~\ref{fig:strategies}A.

The graphs are obtained simulating system (\ref{1Dsys}) for a tent map in
which the ordinate of $B_P$ has been decreased by 2\%, i.e.\ for a system
tolerant to breakpoint misplacements up to 2\%. In Fig.~\ref{fig:pdf-tm}A
it is represented the case in which no misplacement occurs (ideal case),
while Fig.~\ref{fig:pdf-tm}B shows the worst case in which a 2\%
misplacement takes the breakpoint even further down.

By similar considerations, it can be seen that systems based on the
Bernoulli shift suffer from the same inconveniences mentioned
above~\cite{Delgado:El}.

\section{The new approach: systems based on the tailed tent map}
The tailed tent map (Fig.~\ref{fig:maps}B) is a
parametric map, whose secondary breakpoint also defines the position
of the main one, namely

\begin{equation}
\label{eq:tailed}
M(x)=\left\{\hspace{-0.5em}
\begin{array}{ll}
1-x_{b2}+2x & x\in[0,x_{b2}/2)\\
1+x_{b2}-2x & x\in[x_{b2}/2,x_{b2})\\
1-x & x\in[x_{b2},1]
\end{array}\right.,
\end{equation}
with $1/2\le x_{b2} < 1$.
In the following, it will be shown that system (\ref{1Dsys}) with
$I=[0,1]$ and $M$ given by (\ref{eq:tailed}) 
has the same basic properties as the one based on the tent
map, yet is  more robust to implementation errors, whenever a uniform PDF
is the desired goal.

\subsection{Basic properties}
A~ piecewise linear map \mbox{$M:I\rightarrow I$} is also a Markov map if
it takes partition points into partition points, namely a partition
${\mathcal W}=\{(a_0,a_1),\ldots,(a_{N-1},a_{N})\}$, $a_0=a$, $a_{N}=b$,
exists such that if $Q=\{a_0,\ldots,a_N\}$ then $M(Q)\subseteq Q$.  It is
easy to show that whenever $x_{b2}$ is a rational number,
(\ref{eq:tailed}) is a Markov map. Moreover, in this case, $M$ satisfies
also the conditions $1-4$ of Definition~3 in~\cite{Boyarsky:TAMS-255-243}
which assure the PDF unicity and the ergodicity of the map.  Thus, it is
easy to prove that
\begin{equation}
\label{eq:tailedPDF}
\rho(x)=\chi_{[0,1]}(x)
\end{equation} 
is the invariant PDF.  In fact, it can be easily seen that as any $x \in
I$ has a finite number of counterimages, then ${\mathcal P}_{M}$ can be
also written as
\begin{equation}
\label{eq:perron2} 
{\mathcal P}_{M}[\rho](x)=\sum_{M(y)=x}\frac{\rho(y)}{|M'(y)|}.
\end{equation}
Thus, by substituting (\ref{eq:tailedPDF}) in (\ref{eq:perron2}) one
immediately gets 
\[
{\mathcal P}_{M}[\chi_{[0,1]}](x)=
1\cdot\chi_{[0,1-x_{b2}]}(x)+\left(\frac{1}{2}+
\frac{1}{2}\right)\cdot
\chi_{[1-x_{b2},1]}(x).
\]

Knowing the invariant PDF, the Lyapunov exponent for the map can be
computed as \cite{Ott:CDS}
\begin{equation}
\label{eq:Liap-series}
h=\int_{0}^{1}\ln{|M'(x)|}\rho(x)dx,
\end{equation}
which leads to $h=x_{b2} \ln{2}$, proving system chaoticity.  

\subsection{Inherent robustness of the map}
\begin{figure}[ht]
\begin{minipage}{0.42\columnwidth}
\begin{center}
\resizebox{\linewidth}{!}{\includegraphics{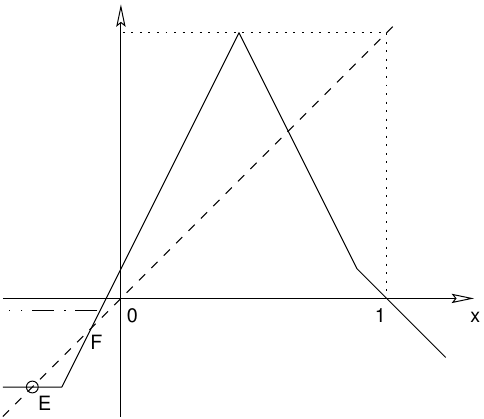}}\par
\centerline{\sffamily\small(A)}
\end{center}
\end{minipage}
\hspace{0.2cm}
\begin{minipage}{0.42\columnwidth}
\begin{center}
\resizebox{\linewidth}{!}{\includegraphics{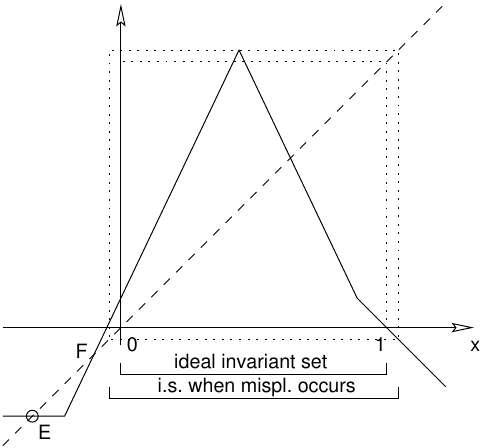}}\par
\centerline{\sffamily\small(B)}
\end{center}
\end{minipage}
\caption{\label{fig:ttm-tol}Graceful alteration of the invariant set
for the tailed tent map.}
\end{figure}

First of all, note that one can design the map in such a way that
point E does not exist, as shown in Fig.~\ref{fig:ttm-tol}A
(dot-dashed line).  Furthermore, even if E exists, the alteration in
the invariant set consequent to breakpoint misplacement is now
\emph{graceful}, i.e.\ the smaller the misplacement the smaller the
invariant set change, as long as this set does not include the
unstable equilibrium point $F$. Therefore, up to a certain
misplacement the invariant set can grow without including E
(Fig.~\ref{fig:ttm-tol}B). More precisely, the lower $x_{b2}$, the
higher the system tolerance. 
As a consequence, a system
based on the tailed tent map can be designed without having to
compromise on the PDF \emph{in principle}. 

\begin{figure}[ht]
\vspace{1ex}
\begin{minipage}{0.45\columnwidth}
\begin{center}
\resizebox{\linewidth}{0.8\linewidth}{\includegraphics{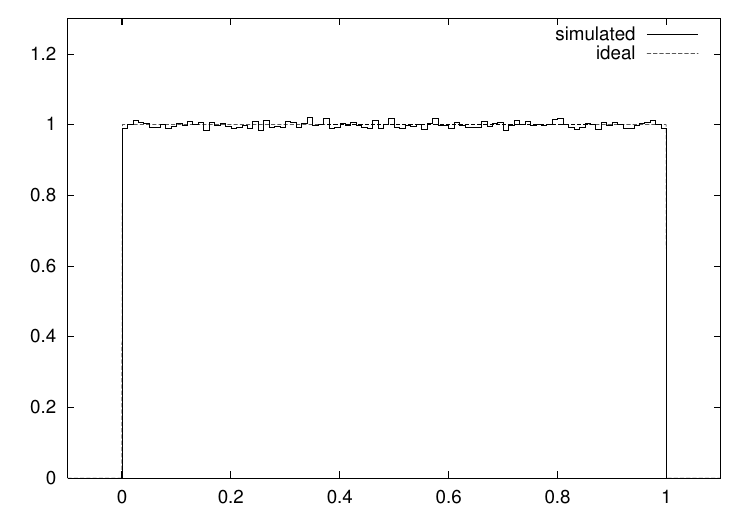}}\par
\vspace{-0.8ex}\centerline{\sffamily\small(A)}
\end{center}
\end{minipage}
\hspace{0.2cm}
\begin{minipage}{0.45\columnwidth}
\begin{center}
\resizebox{\linewidth}{0.8\linewidth}{\includegraphics{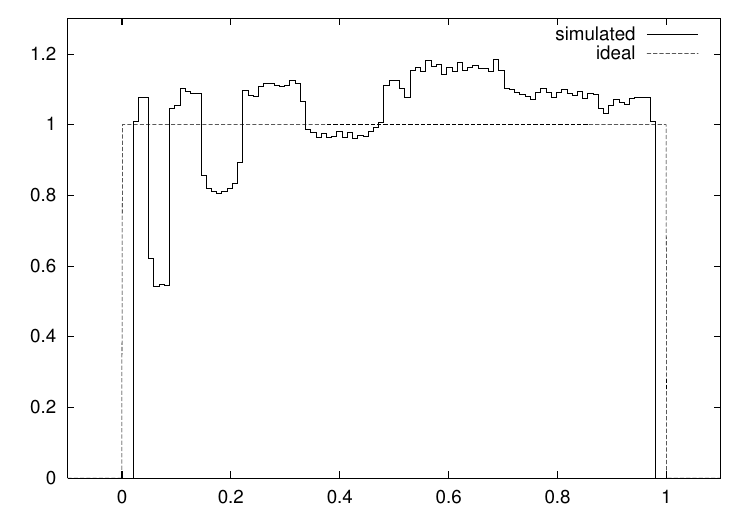}}\par
\vspace{-0.8ex}\centerline{\sffamily\small(B)}
\end{center}
\end{minipage}
\caption{\label{fig:examples}
PDF degradation due to breakpoint misplacement for the tailed
tent map.}
\end{figure}

The effect of implementation errors is shown in Fig.~\ref{fig:examples}.
In particular, Fig.~\ref{fig:examples}A is the ideal case (i.e.\ the map
is implemented correctly) and Fig.~\ref{fig:examples}B represents the
case of a 2\% $B_{P1}$ misplacement. The superiority of the case in
Fig.~\ref{fig:examples}A over the one in Fig.~\ref{fig:pdf-tm}A is
evident and also the case in Fig.~\ref{fig:examples}B is more similar
to an ideal uniform PDF than the one in Fig.~\ref{fig:pdf-tm}B, since
the latter even presents a much larger range of values characterized
by null probability density. The performance improvement becomes even
more visible when one considers probability distributions rather than
densities.

\subsection{Circuit CMOS implementation and performances} 
\begin{figure}[ht]
\vspace{1ex}
\centerline{%
\resizebox{0.84\columnwidth}{!}{
\scalebox{1}[1]{\includegraphics{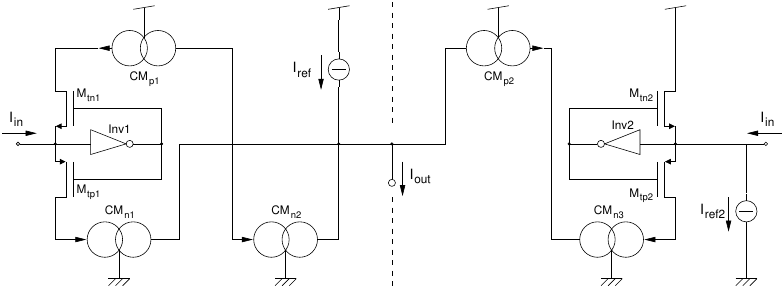}}}}
\caption{\label{fig:ttm-circuit} A circuit for the tailed tent map.}
\end{figure}

Fig.~\ref{fig:ttm-circuit} shows a schematic of the proposed
circuit implementation for the tailed tent map. Its left side is
the same as in Fig.~\ref{fig:tm-circuit}, while it right side
allows  to obtain the  slope
variation necessary to carry out the ``tail.''
Assuming that current mirrors $CM_{p2}$ and $CM_{n3}$
have a $1:1$ ratio, the output current is
$I_{out}=I_{ref}-2|I_{in}|+\mbox{rect}(I_{in}-I_{ref2})$, where
$\mbox{rect}(x)=1/2(x+|x|)$,
namely a tailed tent map apart from the usual change of system reference.

Note that the extra silicon area requested to generate the tailed tent
map is only a small part of the area used for the complete circuit. In
fact, this includes two sample and hold circuits which use a large
proportion of the total area.

\begin{figure}[ht]
\vspace{1ex}
\begin{minipage}{0.45\columnwidth}
\begin{center}
\resizebox{\linewidth}{0.8\linewidth}{\includegraphics{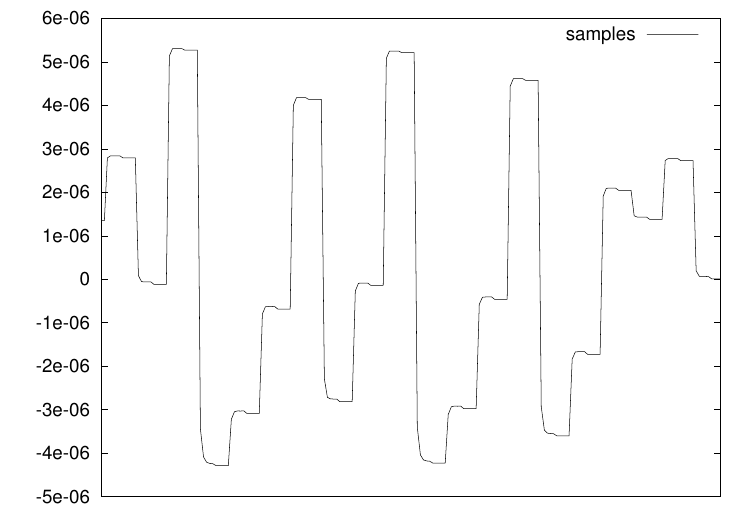}}%
\par
\vspace{-0.8ex}\centerline{\sffamily\small(A)}
\end{center}
\end{minipage}
\hspace{0.2cm}
\begin{minipage}{0.45\columnwidth}
\begin{center}
\resizebox{\linewidth}{0.8\linewidth}{\includegraphics{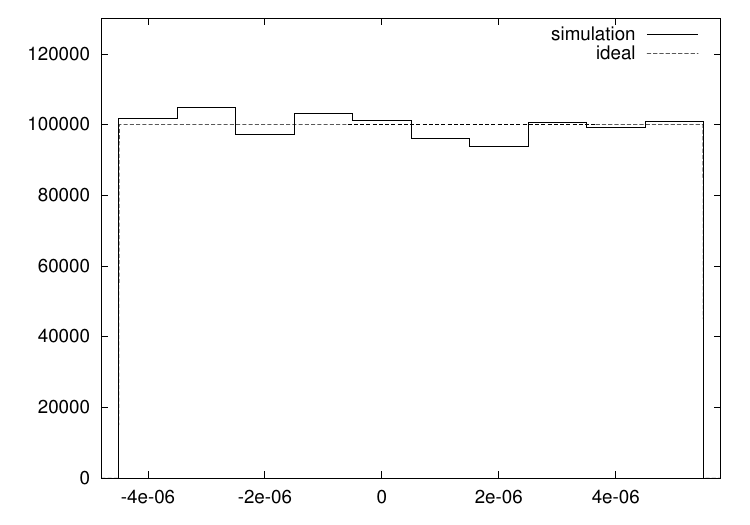}}\par
\vspace{-0.8ex}\centerline{\sffamily\small(B)}
\end{center}
\end{minipage}
\caption{\label{fig:spice} A sample sequence (20 samples) and the PDF
obtained through HSPICE simulations and post-processing (50K
samples used for post-processing).}
\end{figure}

Fig.~\ref{fig:spice} shows the results of device level simulations for the
circuit in Fig.~\ref{fig:ttm-circuit} designed using a standard $1\mu m$
CMOS technology.  To obtain high linearity for the map branches, high
swing cascode mirrors have been used while, to improve maximum frequency
operation, a careful design of the sample and hold stage has also been
considered. Fig.~\ref{fig:spice}A shows a typical chaotic waveform for
$I_{out}(k)$. Fig.~\ref{fig:spice}B represents the approximation of the
PDF obtained by a post-processing executed on the transient analysis
output, for a 750 KHz operating frequency. Deviation from the ideal line
is not only due to circuit misbehavior, but also to the finite number of
samples used to extract the PDF.

In conclusion, the new proposed solution shows several improvements
compared  to previously published results, with respect to the
achievable PDF~\cite{Bean:ISCAS94-6-125} and parameter deviation
robustness~\cite{Delgado:El}.

\vspace{-1ex}
\section*{Acknowledgments}{\small
Sergio Callegari would like to thank Prof. P. U. Calzolari,
Prof. G. Masetti and the EU Erasmus Project for
allowing his research period at King's College, London.}

\bibliographystyle{unsrt}
\bibliography{chaotic,pram,tent,analog}
\end{document}